# The Automated Reaction-Pathway Search reveals the Energetically Favorable Synthesis of Interstellar $CH_3OCH_3$ and $HCOOCH_3$

Yu Komatsu*,†,‡ and Kenji Furuya‡

†*Astrobiology Center, 2-21-1 Osawa, Mitaka, Tokyo 181-8588, Japan.*
‡*National Astronomical Observatory of Japan, 2-21-1 Osawa, Mitaka, Tokyo 181-8588, Japan.*

E-mail: yu.komatsu@nao.ac.jp

**Abstract**

Recent astronomical observations have shown that interstellar complex organic molecules (COMs) exist even in cold environments (~10 K), while various interstellar COMs have conventionally been detected in the hot gas (≳100 K) in the vicinity of high-mass and low-mass protostars. However, the formation pathway of each interstellar COM remains largely unclear. In this work, we demonstrate that an automated reaction path search based on transition state theory, which does not require predetermined pathways, is helpful for investigating the formation pathways of interstellar COMs in the gas phase. The exhaustive search within electronic ground states helps elucidate the complex chemical formation pathways of COMs at low temperatures. Here we examine the formation pathways of dimethyl ether ($CH_3OCH_3$) and methyl formate ($HCOOCH_3$), which are often detected in the cold and hot gas of star-forming regions. We have identified a barrierless and exothermic formation path of $CH_3OCH_3$ by reaction between neutral species; $CH_3O + CH_3 \dashrightarrow H_2CO \cdot CH_4 \dashrightarrow CH_3OCH_3$ is the most efficient path in the large chemical network constructed by our automated reaction path search and is comparable with previous studies. For $HCOOCH_3$, we obtain complex pathways initiated from reactions between neutral species; HCOO and $CH_3$ generate $HCOOCH_3$ and its isomers without external energy. However, we also identified the competing reaction branches producing $CO_2$ + $CH_4$ and $CH_3COOH$, which would be more efficient than the formation of $HCOOCH_3$. Then the gas-phase formation of $HCOOCH_3$ through reactions between neutral species would not be efficient compared to the $CH_3OCH_3$ formation.

# 1. Introduction

Astronomical observations have shown that a variety of interstellar complex organic molecules (COMs) exist in star-forming regions, particularly the hot gas (≳100 K) in the vicinity of high-mass and low-mass protostars (hot cores and hot corinos, respectively), where ices on dust grains are thermally desorbed. It has been proposed that chemical reactions on grain surfaces play essential roles in forming COMs; afterward, they are released into a gas phase via thermal desorption in the hot cores/corinos.[1] Among interstellar COMs, dimethyl ether ($CH_3OCH_3$) and methyl formate ($HCOOCH_3$) are typical and abundant species, and both have conventionally been detected in hot cores and corinos. It has been thought that $CH_3OCH_3$ and $HCOOCH_3$ are formed on warm (≳20 K) dust grains via radical-radical reactions, $CH_3O + CH_3 \dashrightarrow CH_3OCH_3$ and $HCO + CH_3O \dashrightarrow HCOOCH_3$, respectively, through the Langmuir-Hinshelwood mechanism, which requires the surface diffusion of the radicals; thus, it is efficient only on warm dust grains (≳20 K). However, recent quantum chemistry calculations have indicated that $CH_3O$ can be slightly less diffusive in $CH_3O + CH_3 \dashrightarrow CH_3OCH_3$ because the O of $CH_3O$ establishes H-bond interactions with surface water molecules, and the competitive H-abstraction to form $H_2CO$ and $CH_4$ becomes significant at low temperatures because of quantum tunneling effects[2]. Conversely, ion-neutral reactions in the hot (>100 K) gas followed by the sublimation of icy species, which are formed by hydrogen addition reactions in the prestellar phases, may contribute to the presence of some COMs in the hot cores/corinos; for example, it has been proposed that the gas formation of $CH_3OCH_3$ is initiated by the reaction between methanol and protonated methanol, $CH_3OH + CH_3OH_2^+ \dashrightarrow (CH_3)_2OH^+ + H_2O$, followed by proton transfer to molecules with higher proton affinity (e.g., ammonia) and/or dissociative electron recombination to form $CH_3OCH_3$,[3] and quantum chemistry calculations have estimated its rate coefficients.[4]

Recently, both COMs have been detected in low-temperature environments (~10 K), e.g., toward prestellar cores[5,6] and in the cold gas of the outer envelopes of low-mass protostellar sources.[7,8] As $CH_3OCH_3$ and $HCOOCH_3$ are thought to be mainly formed on warm dust grains (≳20 K) via the radical-radical reactions and/or via gas-phase reactions in the hot (≳100 K) gas, the detection indicates that these COMs are formed in the cold gas (~10 K) via gas-phase reactions, on cold grain surfaces followed by non-thermal desorption, or the combination of both.[9–11]

In the cold (~10 K) regions, reactions between neutral radicals can contribute to producing COMs in the gas phase. For the neutral gas-phase formation of $CH_3OCH_3$,

$$CH_3O + CH_3 \dashrightarrow CH_3OCH_3 + h\nu \tag{1}$$

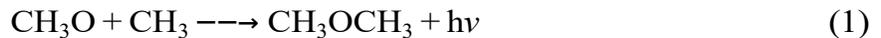

can rapidly occur via radiative association as the efficient radiation of energy from the system.[9] Further, $CH_3O$ and $CH_3$ result in direct H-abstraction.

$$CH_3O + CH_3 \dashrightarrow H_2CO + CH_4, \tag{2}$$

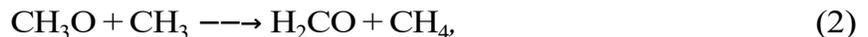

which is also included in astrochemical models.[9,11] Recently, Tennis et al.[12] have estimated the detailed temperature-dependent rate coefficient of the radiative association reaction between $CH_3O$ and $CH_3$ using quantum chemistry calculations. They have shown that although the $CH_3OCH_3$ production via radiative association in the gas phase becomes rapid at low temperatures (10 K), it can not fully explain the high abundance of $CH_3OCH_3$ in cold cores. Therefore, new mechanisms for producing $CH_3OCH_3$ have been anticipated.

In contrast to $CH_3OCH_3$, our knowledge of the interstellar formation mechanisms of methyl formate ($HCOOCH_3$) is considerably limited. Ion-neutral reactions in the gas phase may play an essential role in the interstellar formation of $HCOOCH_3$, similar to $CH_3OCH_3$. For example, methyl cation transfer reaction, $[CH_3OH_2]^+ + HCOOH \longrightarrow [HC(OH)OCH_3]^+ + H_2O$, generates protonated $HCOOCH_3$.[13,14] Further, the production of $HCOOCH_3$ can be assisted by a radiative association via an ion-molecule reaction, such as $CH_3^+ + HCOOH \longrightarrow [HC(OH)OCH_3]^+ + h\nu$ followed by dissociative recombination. Although these ion-neutral reactions were examined based on quantum chemistry calculations and experiments, it cannot explain the observed abundance of $HCOOCH_3$.[13]

In astrochemical models, $HCOOCH_3$ is mainly formed by the neutral gas-phase reaction: $CH_3OCH_2 + O \longrightarrow HCOOCH_3 + H$.[11] $CH_3OCH_2$ is produced by e.g. $CH_3OCH_3 + X \longrightarrow CH_3OCH_2 + X$, where X is a halogen (X = F, Cl). These reactions link between the formations of $CH_3OCH_3$ and $HCOOCH_3$.

In this paper, we present the energetically favorable gas formation of $CH_3OCH_3$ and $HCOOCH_3$, as typical interstellar COMs, via neutral and radical reactions within electronic ground states using an unbiased reaction pathway search method based on the transition theory without specifying the reaction pathways in advance. Recently, we have applied this methodology for the possible interstellar synthesis of nucleobases, aromatic heterocyclic molecules,[15] and we demonstrate here that the approach effectively elucidates and validates the complex chemical networks of COMs, where $CH_3OCH_3$ and $HCOOCH_3$ are smaller molecules than nucleobases. Several formation mechanisms of interstellar $CH_3OCH_3$ are proposed in the previous studies, and they are comparable with our results, whereas the mechanism of HCOOCH$_3$ is not understood, and it is worthwhile to provide a theoretical base with the comprehensive search. Section 2 shows the details of the reaction pathway search calculations. Section 3 presents the chemical reaction pathways for CH$_3$OCH$_3$ and HCOOCH$_3$. The feasibility of their production in star-forming regions is discussed. Finally, Section 4 concludes the findings of the study.

## 2. Methods

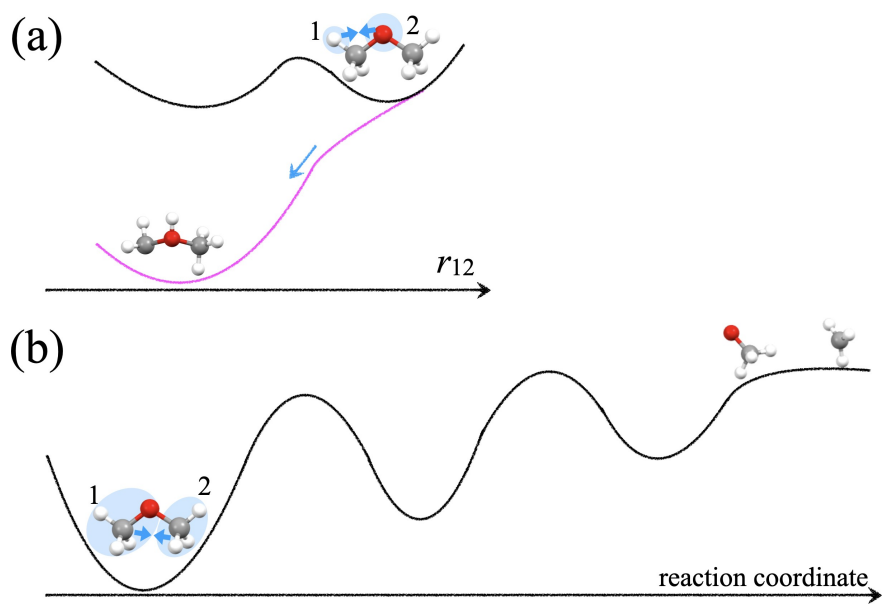

Figure 1: The schematic illustration shows how to search for (A + B → C)-type reactions using the GRRM program[16] in this study. (a) The artificial forces are applied between randomly determined fragments 1 and 2. The updated shape of PES by the force (magenta curve) has no energy barriers, unlike the true PES (black), which easily reaches the local minima. The actual PES can be obtained after the modification of the PES without the force by the subsequent procedure. (b) As a result of multiple attempts as in (a), but starting with different fragments from (a), the dissociation reaction connected to the target molecule on the PES is successfully extracted.

We conducted intrinsic reaction coordinate (IRC) calculations using the global reaction route mapping (GRRM)[16] method to identify the chemical reaction pathways toward tar-get molecules, which are composed of stable structures (EQs), transition states (TSs), and dissociation channels (DCs) on potential energy surfaces (PESs). We primarily focused on the reactions from the DCs to target molecules in the obtained reaction pathways. The artificial force induced reaction (AFIR)[17] method enabled us to effectively extract (A + B → C)-type reactions. As shown in Figure 1, the global reaction network, including dissociation reactions, was obtained using the single-component (SC)-AFIR calculation (SC-AFIR2[18]) as follows. (a) Two fragments in a target (starting) molecule were randomly selected, and artificial forces were applied between fragments 1 and 2. As shown in Figure 1 (a), the original PES $E(r_{12})$ (shown as black curve), where $r_{12}$ was the distance between fragments 1 and 2, was updated as $E(r_{12}) + \alpha r_{12}$ (magenta), where $\alpha$ is the force parameter.[16] As a result of SC-AFIR, no energy barriers (TSs) on the new PES allowed easy access to the local minimum, another stable structure (EQ). The true PESs including TSs were obtained by the subsequent calculations at higher levels of the theory without acting the force. (b) After repeating the SC-AFIR from different fragments and continuing from newly obtained configurations (and subsequent refinement of PESs), the reaction network surrounding the target molecule was constructed. The search was performed within the same chemical formula of a starting molecule, and the isomerization was expected to be obtained as the main reaction, but the dissociation reaction was identified if the distance between the fragments is greater than the threshold value (see below). In the case of (b), the pathway from the newly identified DC to the target starting molecule was extracted. In principle, the search for (A + B → C + D)-type reactions was also possible if the starting configuration consisted of two molecules (C + D). However, only (A + B → C)-type reactions were obtained because we limited one starting molecule (C) in this study.

Thus, we adopted the SC-AFIR calculation by applying two artificial forces between randomly determining fragment molecules with a model collision energy parameter[16] of 1,000 kJ/mol, which corresponds to the upper limit of the energy barrier to overcome. We performed the SC-AFIR calculations at UB3LYP/6-31G(d) level with the "NoBondRearrange" option, limiting the search to local minima with the same bond connectivity to a given configuration. Afterward, to screen unrealistic structures, we reoptimized IRCs with UB3LYP/6-311G(d,p). Moreover, EQs and TSs were reoptimized at the UM06-2X/aug-cc-pVDZ level without re-using the wave functions of the previous procedure as the initial guesses, and the zero-point energy (ZPE) correction was obtained. Finally, we conducted single-point calculations at the UCCSD(T)/aug-cc-pVTZ level without succeeding wave functions from the previous reoptimization of EQs, TSs, and DCs as well. After the procedure, the energy in the obtained pathway was presented in the UCCSD(T)/aug-cc-pVTZ with ZPE. DC was identified when the distance between two fragments was greater than $0.1 \times n \times 2 \times (R_A + R_B)$, where $R_A$ ($R_B$) is the sum of the covalent radii for each atom in fragment A (B). We used $n = 15$, corresponding to a sufficiently large distance between typical neutral hydrogen-bonding fragments. We used the GRRM17 program package[16,19] with Gaussian 16[20] for the quantum chemical calculations.

## 3. Results and discussion

### 3. 1. $CH_3OCH_3$

Through the GRRM calculations, we discovered 22 EQs, 56 TSs, and 1 DC connected to $CH_3OCH_3$ on a PES (see Figure S1 in the Supporting Information). We found an exothermic and barrier-less reaction from the reaction network, $CH_3O + CH_3 \dashrightarrow CH_3OCH_3$, as shown in Figure 2. d-DC ($CH_3O + CH_3$) was obtained as a dissociated molecule. It stabilized in the d-EQ1 configuration ($H_2CO\cdots CH_4$), and the excess energy was sufficiently large to overcome the barriers of the three transition states. All three transition states exhibited similar configurations with similar energies (334.9, 335.2, and 336.0 kJ/mol for d-TS1, d-TS2, and d-TS, whose structures are shown in Supporting Information, respectively), and only the d-TS structure is presented in Figure 2 for the comparison with the previous study at the last paragraph of Section 3.1. After the formation of $CH_3OCH_3$, some transition states were connected to $CH_3OCH_3$ on the PES (not shown). However, none of the states were realized, even when the energy of the d-TS was available in the system without any external energy. Thus, $CH_3OCH_3$ was built as the main product of the reaction with a reaction energy of −370.8 kJ/mol, and the pathway shown in Figure 2 is summarized as follows.

$$CH_3O + CH_3 \dashrightarrow H_2CO\cdots CH_4 \dashrightarrow CH_3OCH_3. \qquad (3)$$

The coordinate of each structure in Figure 1 is provided in the Supporting Information, as one of the key reactions throughout the paper.

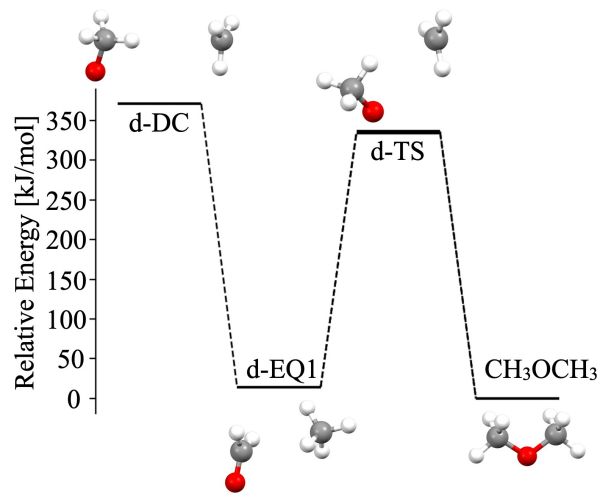

Figure 2: Reaction pathway toward $CH_3OCH_3$ at UCCSD(T)/aug-cc-pVTZ + ZPE. The relative energies to $CH_3OCH_3$ are shown. Black atom: carbon, white: hydrogen, and red: oxygen.

The pathway above was extracted from a part of a circular reaction pathway in a large reaction network, as shown in Figure 3. When either of two more transition states with higher energy than d-TS was realized from d-EQ1, the excess energy was sufficient to form three more transition states from $CH_3OCH_3$, affording another stable molecule d-EQ2, $CH_3O\cdots CH_3$, which is weakly combined by van der Waals interaction. d-EQ2 showed the energy of ∼ 1 kJ/mol, higher than these three transition states. Further, the energy of the transition state between and d-EQ1 ($H_2CO\cdots CH_4$) and d-EQ2 ($CH_3O\cdots CH_3$) was less than that of $CH_3O\cdots CH_3$; thus d-EQ1 was reproduced. In short, we obtained the cyclic pathway (denoted as path $F$), which is as follows.

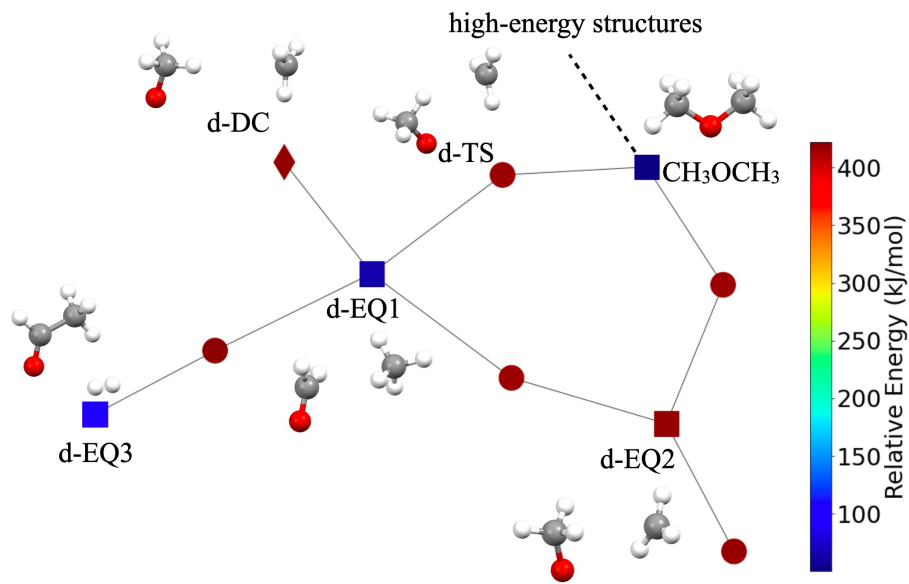

Figure 3: The reaction network around $CH_3OCH_3$ (The same result in Figure 2). Color for each node shows the relative energy. Squared node: EQ, circle: TS, and diamond: DC.

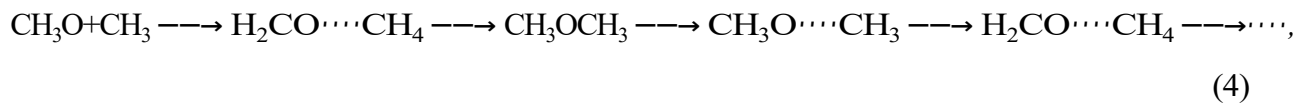

$$CH_3O + CH_3 \dashrightarrow H_2CO\cdots CH_4 \dashrightarrow CH_3OCH_3 \dashrightarrow CH_3O\cdots CH_3 \dashrightarrow H_2CO\cdots CH_4 \dashrightarrow \cdots, \tag{4}$$

where ~ 1 kJ/mol is additionally required to pass $CH_3OCH_3 \dashrightarrow CH_3O\;CH_3$. When the excess energy was insufficient to proceed to the next step, the reaction would be terminated, e.g., by the energy loss due to the emitting photons in the gas phase (radiative association). After that, the pathway shown in Figure 2 is extracted as an effective pathway.

Furthermore, the reverse pathway (path $R$ for the reverse rotation against $F$ for forward) would occur.

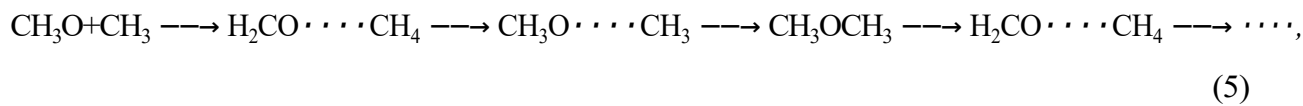

$$CH_3O + CH_3 \dashrightarrow H_2CO\cdots CH_4 \dashrightarrow CH_3O\cdots CH_3 \dashrightarrow CH_3OCH_3 \dashrightarrow H_2CO\cdots CH_4 \dashrightarrow \cdots, \tag{5}$$

To proceed with the reaction $H_2CO\cdots CH_4 \dashrightarrow CH_3O\cdots CH_3$, an activation barrier of 0.2

kJ/mol should be overcome. In both directions of rotation described above, after excess energy was consumed, $CH_3OCH_3$ and $H_2CO + CH_4$ were produced as considerably stable molecules. Although $CH_3CHO\cdots H_2$ (d-EQ3) would build to an extent, the energy was larger than those of d-EQ1 and d-EQ2.

Methoxy ($CH_3O$) radicals have been detected in cold dark clouds along with $CH_3OCH_3$.[5] Conversely, methyl ($CH_3$) radicals have not been detected in cold dark clouds but have been detected toward the galactic center.[21] In gas-phase, $CH_3$ is produced by $CH_3O + H \dashrightarrow CH_3 + OH$ in astrochemical models.[9] When $CH_3$ encountered $CH_3O$ in the cold gas, the reaction pathway in Eq. (3) would contribute significantly to the production of $CH_3OCH_3$. Radiative association, as Eq. (1), could occur in the middle of the reactions. The pathway shown in Figures 2 and 3, as well as the reactions in Eq. (3) – (5), serves as a basis for considering how much $CH_3OCH_3$ is generated.

The reaction pathway in Figures 2 and 3 was partially shown as a roaming radical reaction process in the thermal decomposition of $CH_3OCH_3$ with CCSD(T) calculations,[22] and our results provide a more comprehensive view of the reaction network around $CH_3OCH_3$. $CH_3O + CH_3 \dashrightarrow CH_3OCH_3$ released ∼ 347 kJ/mol of energy, and the conversion from $H_2CO + CH_4$ to $CH_3OCH_3$ required ∼ 343 kJ/mol in their calculations, which were consistent with our results shown in Figure 2. Besides, the configuration of d-TS in Figures 2 obtained by our calculations was close to the structure of the corresponding transition state of theirs (Fig. 14 in Sivaramakrishnan et al. 2011[22]). Thus, although we extracted the chemical reaction pathway shown in Figures 2 and 3 from the quantum chemical search on PES without any assumptions for the path, the obtained pathway was complementary to the previous quantum chemical study and was helpful in speculating the potential reactions in interstellar conditions from the complex reaction network.

## 3.2. HCOOCH$_3$

We found 39 EQs, 61 TSs, and 5 DCs linked to (*cis*-)HCOOCH$_3$ by another GRRM calculation (see also Figure S2 in the Supporting Information). The formation pathways that occurred considerably easily are presented in Figures 4, 5 which are more complicated than those for CH$_3$OCH$_3$. The structures of the transition states in Figures 4 and 5 are shown in Figure 6. The two exothermic pathways shown in Figures 4 and 5 produced HCOOCH$_3$ as an intermediate with some branches. The reaction between HCOO (formyloxy radical) and CH$_3$ shown in Figure 4 proceeded without any external energy (named path I), while one between HCOOH (formic acid) and CH$_2$ (path II), as shown in Figure 5, required 2.4 kJ/mol to form HCOOCH$_3$. Figure 7 shows the hydrogenation of the precursor molecules producing HCOOCH$_3$.

As the transition state between *cis*-HCOOCH$_3$ and *trans*-HCOOCH$_3$, by the GRRM calculation above, we obtained a van der Waals molecule, CH$_3$O···HCO, whose activation energy barrier was huge (382.5 kJ/mol for *trans* → *cis*), instead of a stable transition state without cleavage. We conducted a double sphere (DS)-AFIR calculation[16] for further exploration, which showed a single reaction path between two geometries. DS-AFIR calculations were performed at the UB3LYP/6-31G(d) level, followed by more accurate evaluations described in Section 2. As expected, we found a more stable transition state for a single molecule and evaluated the energy at UCCSD(T)/aug-cc-pVTZ + ZPE. The activation barriers were 52.06 kJ/mol for *cis* → *trans* and 32.17 kJ/mol for *trans* → *cis*, which essentially corresponded with a previous quantum calculation with MP2/6-31++G(d,p) level.[23] Thus, we focused only on *cis*-HCOOCH$_3$ hereafter because the conversion required enormous energy and was not plausible at a temperature less than 30 K.

The starting molecules in the pathway shown in Figure 4 (path I) were HCOO and CH$_3$ (f1-DC). These molecules produced f1-EQ3, f1-EQ4, f1-EQ5, f1-EQ7, and f1-EQ8. We summarized the expected reactions as follows.

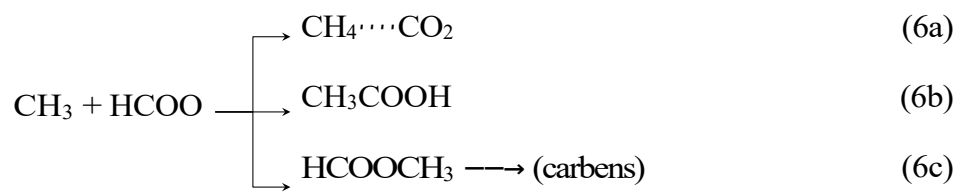

$$\text{CH}_3 + \text{HCOO} \longrightarrow \begin{cases} \text{CH}_4\cdots\text{CO}_2 & (6a) \\ \text{CH}_3\text{COOH} & (6b) \\ \text{HCOOCH}_3 \dashrightarrow (\text{carbens}) & (6c) \end{cases}$$

Radical f1-DC was stabilized toward f1-EQ1, $CO_2$ + $CH_4$. f1-EQ1 was converted to f1-EQ3, which had a different configuration of $CO_2$ + $CH_4$, with a small energy barrier of the transition state f1-TS2 (0.4 kJ/mol). The energy levels for f1-EQ1, f1-TS-2, and f1-EQ3 were almost identical. This path in Eq. (6a) was one of the most plausible reactions from HCOO + $CH_3$. However, the excess energy would conquer two more transition states, f1-TS1, and f1-TS3. From f1-TS1, $CH_3COOH$ (acetic acid) was produced as shown in Eq. (6b). f1-TS1 formed f1-EQ2, and f1-EQ4 ($CH_3COOH$) was produced after the formation of f1-TS4. f1-TS3 in another branch produced $HCOOCH_3$ as displayed in Eq. (6c). And then, $HCOOCH_3$ had two more branches of the reaction pathway with higher energy than that of $HCOOCH_3$ but accessible due to the excess energy. The production of $HCOOCH_3$ was highly exothermic (395.8 kJ/mol). f1-TS5 formed f1-EQ5, whereas f1-TS6 was stabilized in f1-EQ6. f1-EQ6 was converted to f1-EQ7 and f1-EQ8, over f1-TS7 and f1-TS8, respectively. However, these isomers of $HCOOCH_3$ (f1-EQ4, f1-EQ5, f1-EQ7, and f1-EQ8) would not be realized as final products since they were unstable carbens.

$CH_3$ is an abundant radical, as discussed above. The formyloxy radical (HCOO) is an isomer of a more stable COOH radical, and both are intermediates in CO + OH $\dashrightarrow$ $CO_2$ + H.[24,25] As discussed so far, the pathway in Figure 4 starting from $CH_3$ and HCOO was the most efficient for the formation of $HCOOCH_3$ in the obtained chemical network, as a result of the thorough search within the system of $C_2H_4O_2$ in ground states. Radiative association could promote the formation of $HCOOCH_3$ in the gas phase as discussed for $CH_3OCH_3$ to some extent. If the pathway is effective as an interstellar reaction, $HCOOCH_3$ and CH$_3$COOH are given as intermediates in the pathway whose main products are CH$_4$ and CO$_2$, and then the yield of HCOOCH$_3$ might be small. In this case, non-thermal desorption of HCOOCH$_3$ upon the formation on the grain surface may be more efficient than the gas-phase formation pathway. Indeed, recent laboratory studies have found that the photolysis of CH$_3$OH in water ice at 10 K leads to the formation of HCOOCH$_3$.[26]

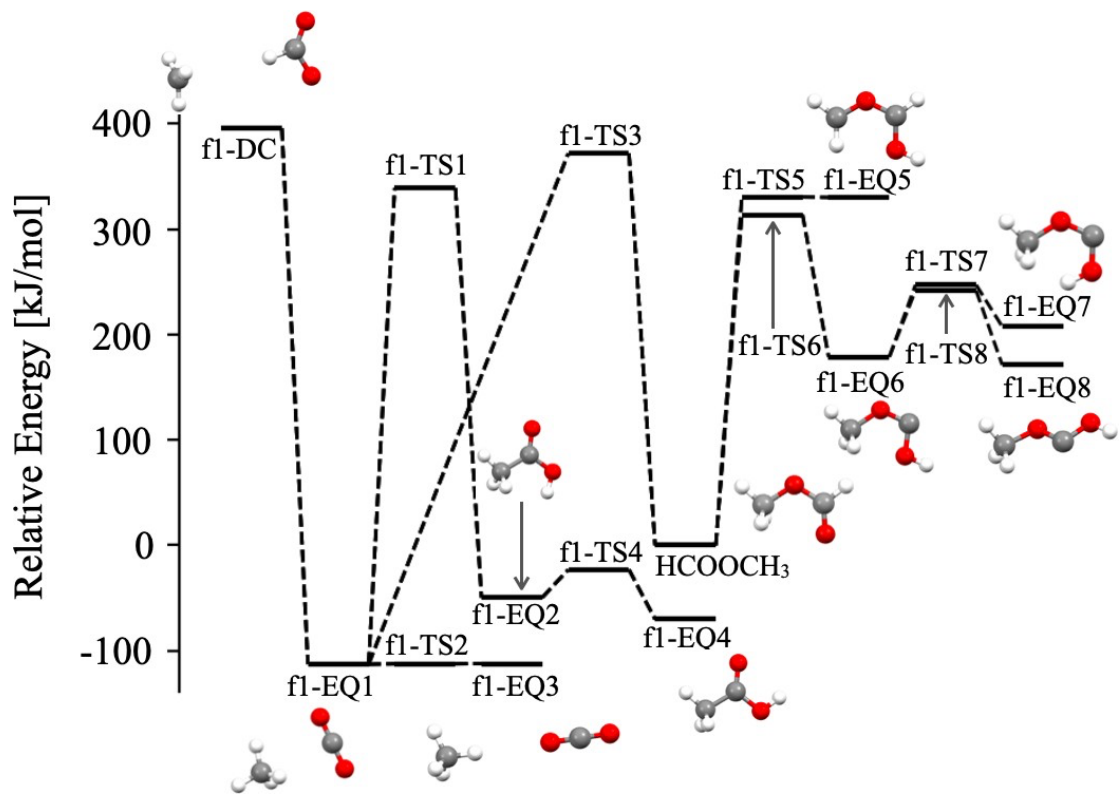

Figure 4: Reaction pathway (path I) including HCOOCH$_3$ at UCCSD(T)/aug-cc-pVTZ + ZPE.

There is another pathway for producing HCOOCH$_3$ (path II), as shown in Figure 5. For the reaction to progress, 2.4 kJ/mol was required in the middle of the reaction path (f2-EQ2 → f2-TS3). However, HCOOCH$_3$ was considerably stable in the reaction pathway, unlike the path in Figure 4, after overcoming the barrier. The dissociated molecules, HCOOH + CH$_2$ (f2-DC), became stable in the f2-EQ1 configuration, and two transition states (f2-TS1 and f2-TS2) were readily realized owing to the excess energy. f2-TS1 produced f2-EQ3 and f2-EQ5, biradicals, via f2-TS4. This reaction proceeded without external energy. As another route, f2-EQ2 was converted from f2-TS2, and it required 2.4 kJ/mol to overcome the barrier of f2-TS3 and produce f2-EQ4. f2-TS-5 was easily realized and then stabilized as HCOOCH$_3$ (−404.9 kJ/mol from the reactant). Subsequently, the reactions leading carbens in Figure 4 existed. Accordingly, HCOOCH$_3$ would be produced to an extent when sufficient external energy is gained in the system because HCOOCH$_3$ is stable significantly compared with the other configurations. Otherwise, biradicals f2-EQ3 and f2-EQ5 would be the leading products in the reaction.

The formic acid (HCOOH) has been detected in various interstellar environments, and the hydrogenation of the HO–CO complex as a surface reaction is suggested to produce HCOOH at low temperatures effectively.[27] As another reactant, methylene radical (CH$_2$) is abundant in interstellar COMs. Thus, these reactants would be available, and the pathway shown in Figure 5 could be possible in hot cores/corinos (to overcome the 2.4 kJ/mol barrier) to form HCOOCH$_3$. However, it is unlikely to occur in the interstellar cold molecular medium. As an interstellar synthesis of HCOOCH$_3$, reactions between two formaldehyde molecules (2HCHO −−→ HCOOCH$_3$) were proposed.[28] Nevertheless, this would be less efficient than reactions in Figures 4 and 5 because the barriers in MP2/6–311G(d,p) are huge (49.94 kcal/mol and 50.76 kcal/mol in the gas phase and PCM, respectively).

Figure 7 presents three different configurations of hydrogenation, and each configuration directly formed HCOOCH$_3$ without activation energy barriers. However, considerable energy due to hydrogenation caused numerous reactions, not only producing HCOOCH$_3$ (See Figure S2 in Supporting Information). Resultantly, varieties of molecules, including HCOOCH$_3$, would be produced when such reactions are dominant in a system. The hydrogenation process would be significant in forming COMs on the grain surface at lower temperatures, as described in the Introduction. Though our calculations did not explicitly deal with the grain, if the reaction works, two CO molecules on the grain could be protonated one by one, and then, HCOOCH$_3$ was finally produced from the configurations in Figure 7. Or, radiative association between a hydrogen and COMs in the gas phase could build HCOOCH$_3$ to some extent.

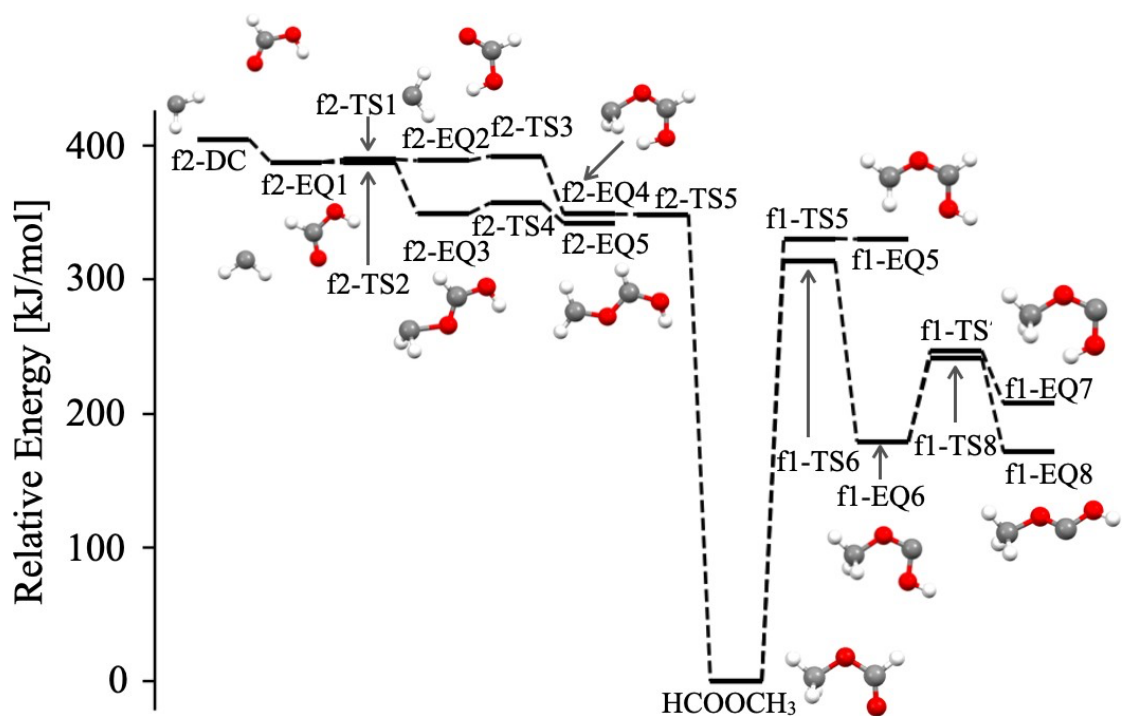

Figure 5: Another reaction pathway (path II) including HCOOCH$_3$ at UCCSD(T)/aug-cc-pVTZ + ZPE.

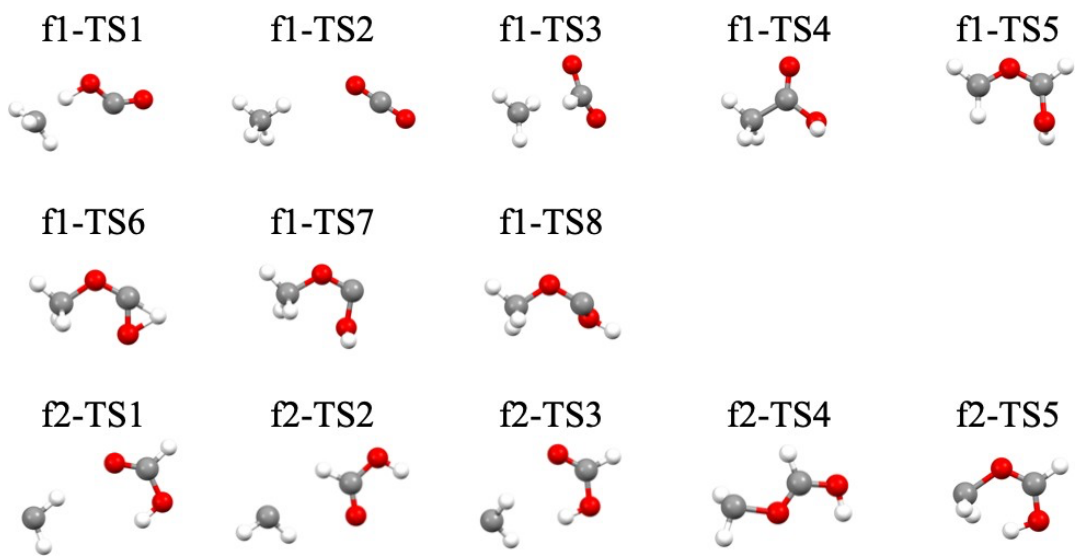

Figure 6: The structures of the transition states in Figures 4 and 5.

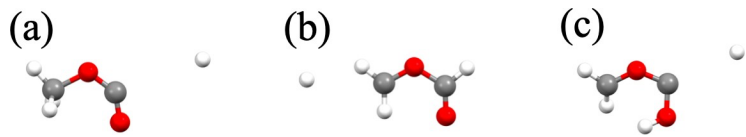

Figure 7: Three configurations of the hydrogenation toward HCOOCH$_3$.

## 3.3. Other reactions

So far, all (A + B → C)-type reactions in the obtained chemical network have been discussed. However, the obtained network was mainly composed of EQs and TSs, i.e., isomerization. Here, the (A → B)-type reactions obtained in this work were summarized in Table 1 in addition to the bimolecular reactions (A + B → C). The energy barriers and minimum steps to the target molecules obtained from GRRM calculations are presented. The barriers were re-evaluated using DS-AFIR as described in the first paragraph of Section 3.2 for several reactions like *cis−* and *trans−* conversion of HCOOCH$_3$.

Ethanol (C$_2$H$_5$OH) is an isomer of CH$_3$OCH$_3$ (DME), and it has been frequently detected in star-forming regions.[29] HCOOCH$_3$ (MF) has two isomers, glycolaldehyde (HCOCH$_2$OH, denoted as GA) and acetic acid (CH$_3$COOH, AA). GA was detected around the galactic center and the Sagittarius B2 (Sgr B2) (N) dust cloud,[28,30] and AA was detected for the first time in the hot molecular core Sgr B2 (N- LMH)[31] and then in a variety of regions.[28] Compared with the biomolecular reactions that we discussed so far (paths *F* and *R* for DME, and paths I and II for MF), the barriers of isomerization were quite high. Nevertheless, we have summarized these reactions for the sake of completeness.

Table 1: The summary of the obtained formation pathways of $CH_3OCH_3$ (DME) and $HCOOCH_3$ (MF). The maximum energy barriers and minimum steps to form the product from the reactant were shown for each path. The energy barriers shown in brackets were obtained by further estimation using DS-AFIR. "ND" indicates that the transition state was not found by DS-AFIR calculations.

*The values for the shortest path from the reactant to the product in the cyclic pathway are shown.

| DME | $F*$ (Eq. 4) | $R*$ (Eq. 5) | Ethanol | | |
|---|---|---|---|---|---|
| reaction | $CH_3O+CH_3$ $\rightarrow$ DME | $CH_3O+CH_3$ $\rightarrow$ DME | $C_2H_5OH$ $\rightarrow$ DME | | |
| barrier [kJ/mol] | 0.0 | 0.2 | ~553 (ND) | | |
| minimum steps | 2 | 3 | 1 | | |
| MF | I (Fig. 4) | II (Fig. 5) | $trans \rightarrow cis$ | GlycolAldehyde | Acetic Acid |
| reaction | $CH_3+HCOO$ $\rightarrow$ MF | $CH_2+HCOOH$ $\rightarrow$ MF | $trans-$MF $\rightarrow$ (cis-)MF | GA $\rightarrow$ MF | AA $\rightarrow$ MF |
| barrier [kJ/mol] | 0.0 | 2.4 | 382.5 (32.17) | ~430 (468.93) | ~420 |
| minimum steps | 2 | 4 | 1 | 1 | 2 |

## 4. Conclusion

We aimed to elucidate the complex reaction networks toward $CH_3OCH_3$ and $HCOOCH_3$ and extract pathways that would proceed in star-forming regions using GRRM calculations. For $CH_3OCH_3$, an exothermic and barrier-less reaction pathway was obtained. $CH_3O + CH_3 \rightarrow H_2CO + CH_4 \rightarrow CH_3OCH_3$ was the most effective from a complex reaction pathway including circular reactions (paths $F$ and $R$). The obtained pathway would be helpful to consider the possible interstellar formation of $CH_3OCH_3$ in the gas phase, and it was consistent with the previous study on the roaming reaction pathway in the thermal decomposition of $CH_3OCH_3$.[22]

On the other hand, the pathways obtained for $HCOOCH_3$ were more complicated than those for $CH_3OCH_3$. Formyloxy (HCOO) and methyl ($CH_3$) radicals formed $HCOOCH_3$ through a barrier-less and exothermic pathway (path I). $HCOOCH_3$ and acetic acid would be generated through this reaction, although $CH_4$ and $CO_2$ were the main products. Another pathway, path II, started from the formic acid and $CH_2$ with a 2.4 kJ/mol activation barrier, which could be effective in hot cores/corinos. Thus, these pathways could not be prominent for the interstellar formation of HCOOCH$_3$ considering they had more stable intermediates or required external energy.

Further validation of possible interstellar reactions at higher levels of theory or laboratory experiments based on our chemical paths would be meaningful to extract effective reactions from complicated reaction networks. As a significant extension of this study, the estimation of abundances of interstellar molecules by astronomical models based on Arrhenius-type equations would be crucial for comparison with observations, using the rate coefficients of key reactions obtained from GRRM calculations. Moreover, in this study, we focused on neutral reactions; however, the methodology applied to ion-neutral reactions may effectively elucidate the formation of interstellar COMs.

## Acknowledgement

We thank an anonymous reviewer for comments to help us improve the manuscript. This work was in part supported by the Astrobiology Center Program of National Institutes of Natural Sciences (NINS) (Grant Number AB021003).

## Supporting Information Available

The chemical networks of CH$_3$OCH$_3$ and HCOOCH$_3$ obtained by the GRRM calculation and a part of XYZ information in the figures are provided.

# Supporting Information:

# The Automated Reaction-Pathway Search reveals the Energetically Favorable Synthesis of Interstellar $CH_3OCH_3$ and $HCOOCH_3$

Yu Komatsu*,†,‡ and Kenji Furuya‡

†*Astrobiology Center, Osawa 2-21-1, Mitaka, Tokyo, Japan.*
‡*National Astronomical Observatory of Japan, Osawa 2-21-1, Mitaka, Tokyo, Japan.*

E-mail: yu.komatsu@nao.ac.jp

# Conformations in Figure 2

== d-DC ==

O -2.085232839723 1.764390477325 -0.325045486538

C 2.483415315622 -1.145878873061 0.001046267043

C -1.978614991411 0.448146130126 0.041900104986

H 2.549218402119 -1.160652406879 1.084048072172

H 2.671776278927 -0.224894271820 -0.540760136720

H 2.291863483663 -2.065264747702 -0.542560800593

H -2.442433575172 0.236436601130 1.018113704959

H -0.939594909222 0.081791861789 -0.009816151860

H -2.550395164805 -0.112009770911 -0.726877573457

== d-EQ1 ==

O -1.126645116051 0.969021335039 1.481624641823

C 1.563610279117 -0.684167872187 -0.671908455634

C -1.249699663194 0.193752567460 0.571076764524

H 1.454389986284 0.400691657680 -0.769876803190

H 2.521338427855 -0.992449750584 -1.101062240189

H 1.535485550795 -0.958526405744 0.387521015349

H -1.265567333799 -0.903673411033 0.738375039557

H 0.751660843430 -1.188468747566 -1.207354920751

H -1.354078285800 0.539736380252 -0.478668212197

== d-TS1 ==

O -1.521495069544 1.561448226594 -0.113488793940

C 1.811553795665 -0.851839982802 -0.043187708630

C -1.432361798099 0.204076625900 0.038577677271

H 1.870461253021 -1.294927031709 0.945557309450

H 1.920593402633 0.221197697456 -0.162592768981

H 1.745613177059 -1.489581623057 -0.918586398419

H -0.900163898809 -0.088388624172 0.958586490054

H -1.012248448001 -0.297861739481 -0.848402044659

H -2.481950413896 -0.142058548713 0.143584237867

== d-TS2 ==

O 1.234375546544 1.012458836149 0.700899922671

C 1.564942320724 -0.220820464043 0.219608787759

C -1.859074979614 -0.328615272307 -0.285938655834

H 1.925531972514 -0.904590199565 1.005054421839

H 2.250585930321 -0.178522536159 -0.641809475636

H 0.600383673790 -0.643434648264 -0.152221654321

H -1.737873717274 -0.120949133350 0.772604098306

H -1.618057958077 0.442258183383 -1.011188706169

H -2.360810788943 -1.235719765837 -0.606960738624

== d-TS ==

O -0.814583448518 1.379371861734 -0.355632286634

C 1.745593613328 -0.814808759762 0.416866976956

C -1.480571432151 0.187172552334 -0.349288660391

H 1.560016666032 -0.434176339075 1.416353417416

H 1.759395901163 -0.126614068364 -0.422494598035

H 2.059474456015 -1.844696517092 0.280967087473

H -0.731232441922 -0.549911283789 0.025024729561

H -1.765638217337 -0.145527679218 -1.360566841858

H -2.332453096628 0.171255233232 0.348818175513

== CH$_3$OCH$_3$ ==

O 0.000000107198 0.595397173657 -0.000016201225

C 1.164116444215 -0.194975390594 0.000020870010

C -1.164116012394 -0.194975709887 -0.000012597215

H 1.207470421948 -0.837191211527 0.895956371726

H 2.023221208125 0.482738464979 0.000015846037

H 1.207496170140 -0.837237180302 -0.895880471486

H -1.207495829602 -0.837191114240 0.895921958231

H -1.207469547068 -0.837237940883 -0.895914884812

H -2.023220962562 0.482737908795 -0.000042891262

## Obtained reaction networks

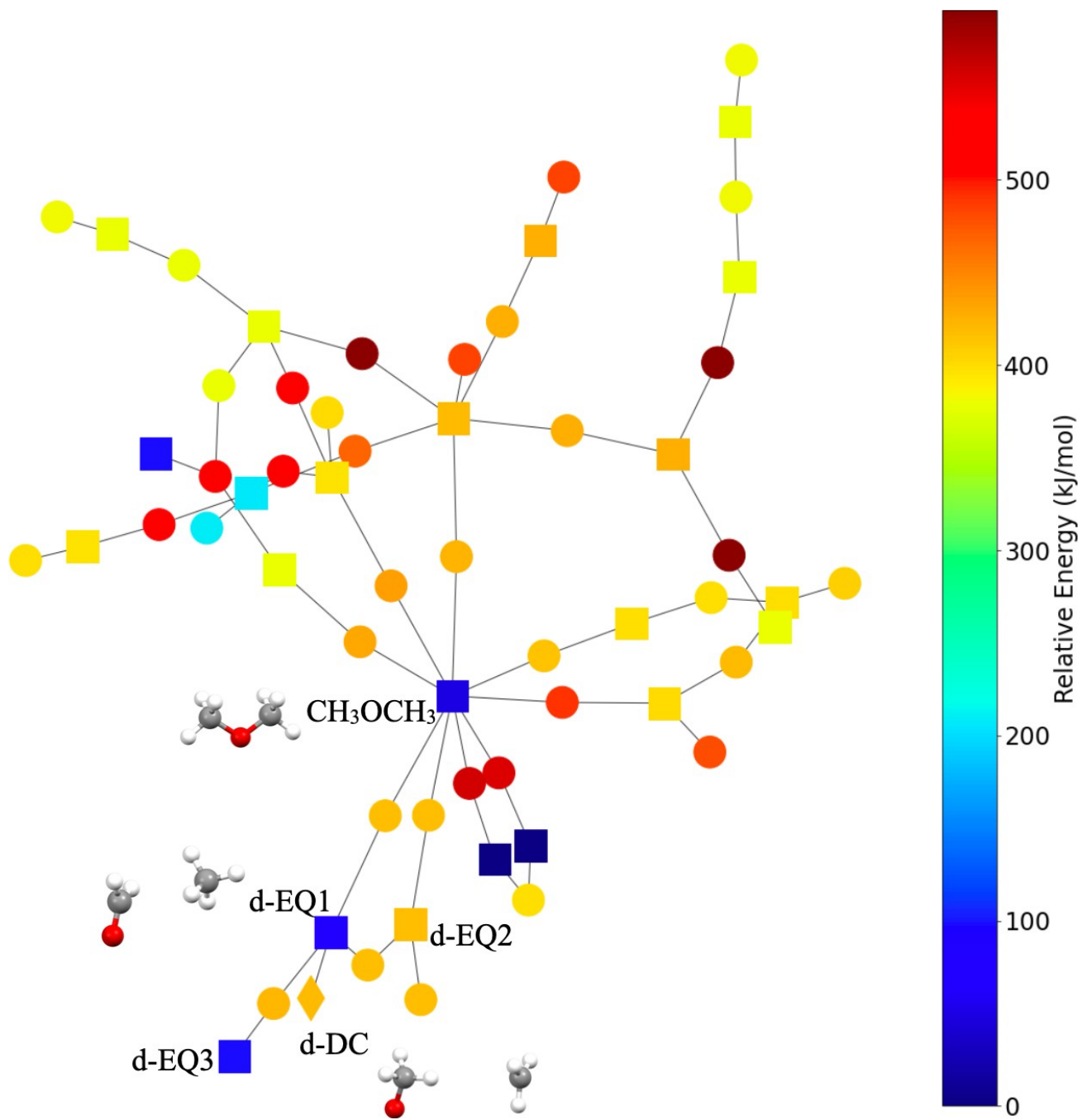

**Figure S1** The chemical network of $CH_3OCH_3$ obtained by the GRRM calculation. The color for each node shows the relative energy at UCCSD(T)/aug-cc-PVTZ level without ZPE correction. Squared node: EQ, circle: TS, and diamond: DC. When a node has multiple TSs, TS with higher energy is not displayed basically, for simplicity.

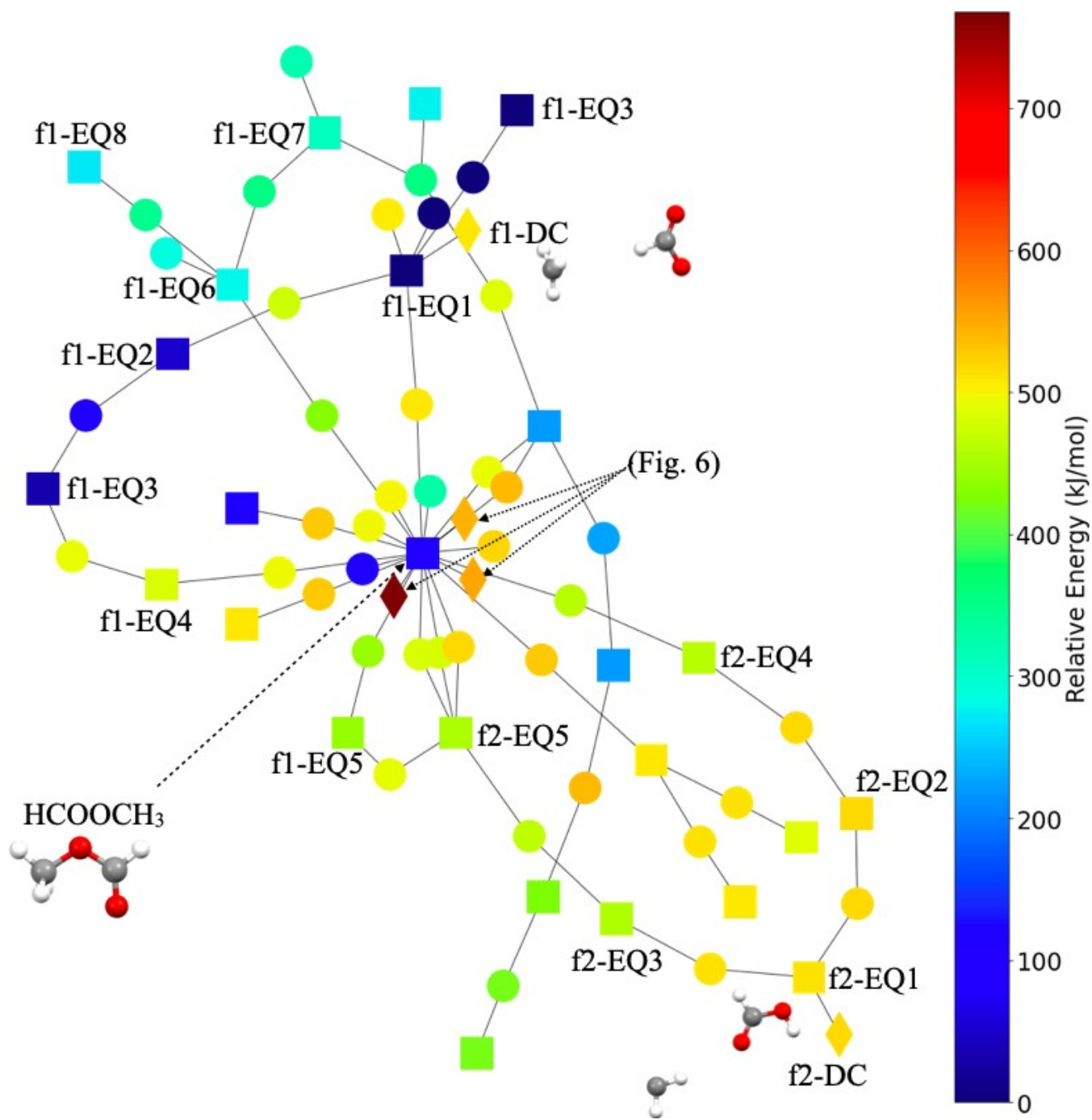

**Figure S2** The chemical network of HCOOCH$_3$ obtained by the GRRM calculation. The EQs, TSs, and DCs are shown only within 520, 540, and 770 kJ/mol, respectively, from the most stable configuration (f1-EQ1).